%% file: zpaper.tex
\documentclass{aa}  

\usepackage{graphicx}
\usepackage{txfonts}
\usepackage{microtype}
\usepackage{hyperref}
\usepackage{url}                    
\usepackage{multirow}               
\usepackage{subcaption}             

\begin{document}

   \title{The Growing Impact of Unintended Starlink Broadband Emission on Radio Astronomy in the SKA-Low Frequency Range}

   \titlerunning{The Growing Impact of Starlink Broadband Unintended Emission}

   \author{D. Grigg
          \inst{1,2},
          S. J. Tingay\inst{1}, and
          M. Sokolowski\inst{1}
          }

    \authorrunning{Grigg et al.}

   \institute{International Centre for Radio Astronomy Research, Curtin University, Bentley, WA, 6102, Australia\\
              \email{dylan.grigg@icrar.org}
         \and
             DUG Technology, 76 Kings Park Rd, West Perth, 6005, WA, Australia}

   \date{Received XXXXXXX XX, XXXX; accepted XXXXXXX XX XXXX}


   \abstract
   {We present the largest survey to date characterising intended and unintended emission from Starlink satellites across the SKA-Low frequency range. This survey analyses $\sim$76 million full sky images captured over $\sim$29 days of observing with an SKA-Low prototype station - the Engineering Development Array 2 - at the site of SKA-Low. We report 112,534 individual detections of 1,806 unique Starlink satellites, some emitting broadband emission and others narrowband emission. Our analysis compares observations across different models of Starlink satellite, with 76\% of all v2-mini Ku and 71\% of all v2-mini Direct to Cell satellites identified. It is shown that in the worst cases, some datasets have a detectable Starlink satellite in $\sim$30\% of all images acquired. Emission from Starlink satellites is detected in primary and secondary frequency ranges protected by the International Telecommunication Union, with 13 satellites identified between 73.00 - 74.60 MHz and 703 identified between 150.05 - 153.00 MHz. We also detect the reflections of terrestrial FM radio off different models of Starlink satellites at 99.70 MHz. The polarisation of the broadband emission shows the flux density of two orthogonal polarisations is anti-correlated with temporally shifting spectral structure observed. We compare our results to previous EDA2 and LOFAR results and provide open public access to our final data products to assist in quantifying future changes in this emission.}


   \maketitle


\section{Introduction}

Mega-constellations of satellites are being deployed to provide global internet coverage and communication at a previously unattainable scale. Several operators such as SpaceX (Starlink constellation; USA), Amazon (Kuiper constellation; USA), and Eutelsat (OneWeb constellation; UK), control the largest existing mega-constellations, with other large mega-constellations planned from Shanghai Spacecom Satellite Technology (G60/Qianfan constellation; China) and the Russian Federation (Sfera constellation; Russia). These operators plan to launch hundreds to tens of thousands of satellites into orbit to provide their services. Low Earth orbit (LEO; typically 400 - 800 km altitude) is desirable for these satellites, as the latency of communications between satellites and the ground stations is minimised compared to geostationary orbit (GEO; $\sim$36,000 km altitude), and requires fewer satellites than would be needed in higher latency orbits for similar network bandwidth.

\input{./results_table.tex}

Whilst the services these companies provide are highly sought-after, there are drawbacks to launching this many satellites into LEO, some of which are only being discovered after the launch of these satellites. Astronomy is one of the domains which suffers.

The effects of satellites interfering with optical astronomy have been well documented \citep{2020AJ....160..226T, 2021arXiv211109735M, 2023AJ....166...59F}, but the detrimental effects on radio astronomy are just being discovered. Satellites are interfering with radio telescopes in the most radio-quiet places on Earth. During the commissioning stages of the Murchison Widefield Array \citep[MWA,][]{mwa1}, located in Western Australia, satellites were detected \citet{offringa2015}, and additional detections were analysed in \citet{steve} and \citet{hen1}. Other examples include the LOFAR telescope in the Netherlands \citep{lofar_passive, lofar, bassa_starlink} and the ALMA telescope in Chile \citep{alma_interference}.

The SKA-Low prototype station, the Engineering Development Array 2 \citep[EDA2,][]{eda2}, has also been used to analyse the effect of satellites on radio astronomy data \citep{tingay_eda_ssa, soko_eda2, grigg, grigg_starlink, grigg_paper3}. The threat satellite interference poses to radio astronomy as the low-frequency half of the Square Kilometre Array (SKA-Low) comes online is important to understand. The EDA2 has the benefit of being at the same location as the SKA-Low in Western Australia. It has the same number of antennas as an SKA-Low station - appropriate for understanding how satellites might affect SKA-Low, in advance of SKA-Low itself.

Satellite downlink frequencies are allocated and regulated by the International Telecommunication Union's Radiocommunications sector (ITU-R\footnote{\url{https://www.itu.int/pub/R-HDB-22-2013}}). A small fraction of the radio spectrum across the SKA-Low's bandwidth has protection for radio astronomy (3.7\%), managed by the ITU-R\footnote{\url{https://www.itu.int/en/ITU-R/information/Pages/default.aspx}}. Although these protections exist, a succession of research has shown that an increasing number of satellites are being detected transmitting unintended electromagnetic radiation (UEMR) outside their designated downlink frequencies, and sometimes at these protected frequencies \citep{tingay_eda_ssa, soko_eda2, grigg, lofar, grigg_starlink, bassa_starlink, grigg_paper3}. The source of this UEMR is likely leakage from some electrical component/s of the satellites. Presently, such emissions are not explicitly covered by ITU regulations, although necessary discussion is currently underway.

The Starlink constellation of satellites is the largest mega-constellation, with over 7,000 satellites currently in orbit. Launches are frequent, and as an indication, 477 Starlink satellites were launched during the four month period of data acquisition for this study.

There have been papers from two main groups investigating the effect of Starlink satellites on radio astronomy. Our team produces full sky images with the EDA2 to detect and identify satellites. Recently, we published work that analysed 20 hours of data at 137.5 MHz and 23 hours at 159.4 MHz and found Starlink satellites transmitting at the two frequencies \citep{grigg_starlink}. At 137.5 MHz, 136 Starlink satellites were detected transmitting intended emission every 100 s, while at 159.4 MHz, a train of Starlink satellites was found to be transmitting UEMR. Communication with SpaceX engineers suggested the UEMR originated from the propulsion/avionics system of the satellites as they were orbit-raising at the time of detection.

Another team has used the Low Frequency Array \citep[LOFAR,][]{lofar_desc} to detect Starlink satellites. For this work, many simultaneous electronically steered beams were formed on the sky and directed at Starlink trains. An initial study using this technique detected 68 Starlink satellites in a one hour observation over the frequencies 110 to 188 MHz \citep{lofar}. In a second paper, \citet{bassa_starlink} detected 29 Starlink satellites in a one hour observation at 10 - 88 MHz and 97 Starlink satellites in another one hour observation at 100 - 188 MHz. Both studies observed the Starlink satellites emitting both narrowband and broadband emission.

Our current study surveys frequencies key to SKA-Low science goals, as well as protected frequencies, using 29 separate 24 hour observations at different frequencies. The benefit of the EDA2 is that no complex beam-forming is required, with the full sky, horizon to horizon, observed throughout the observation. The system can passively search for satellites autonomously in an unbiased manner and generate a list of satellite identifications with little user interaction.

This paper focuses specifically on the detection and characterisation of Starlink satellites, as Starlink satellite numbers vastly outnumber detections of all other satellites combined in our data. The following discussion characterises the detected signals, compares current observations to previous studies, and demonstrates the capability of the EDA2 to identify satellites autonomously.

\section{Observations and data processing}
\label{sec:methods}

The EDA2 is a radio telescope featuring 256 dual polarisation dipoles in a 35 m diameter footprint. It is located at the CSIRO Murchison Radio-astronomy Observatory (Inyarrimanha Ilgari Bundara) in Western Australia, a dedicated radio quiet zone (RQZ) by national Australian legislation \citep{rqz}. The X (East/West) and Y (North/South) polarisations were recorded from each dipole, and all dipole pairs were cross-correlated to form XX and YY polarisation products. The resulting visibilities were averaged over two second periods with 32 frequency channels at a resolution of 28.9 kHz each, totalling 0.926 MHz.

Twenty nine separate observations at 24 unique frequencies were made with the EDA2 for this work. Some frequencies were acquired twice due to a number of small lapses in recording in some datasets. Important frequencies for radio astronomy and others with various levels of legislative protections globally were chosen. These are detailed in Table \ref{table_starlink}. Eleven frequencies were chosen from 100 MHz to 200 MHz every 10 MHz, sampling the approximate frequency range related to Epoch of Reionisation (EoR) science - one of the key science goals for SKA-Low \citep{eor_explanation}. 

Three frequency bands are protected under primary or secondary allocations in the ITU-R which fall within the SKA-Low observing frequency range of 50 - 350 MHz. These are 73.00 - 74.60 MHz, 150.05 - 153.00 MHz, and 322.00 - 328.6 MHz. Seven datasets were chosen to overlap these frequencies to search for satellites transmitting in these allocations.

The frequency 137.5 MHz has been previously analysed in \citet{grigg_starlink}, so two adjacent frequency bands on either side were captured for this work, as they are well populated with satellite downlink frequencies allocated by the ITU. 

\subsection{Data Processing}

The visibility amplitudes and phases were calibrated using the method explained in \citet{soko_eda2} derived from \citet{eda2_old_calibration} with the software package \texttt{Miriad} \citep{miriad}. The flux densities used to calibrate against the Sun were taken from the work by \citet{benz_sun_model}. \texttt{Miriad} was also used to image the visibilities with 100 iterations of CLARK CLEANing applied \citep{clark}. The resulting images were phase-centred to the zenith, of the full sky, and their image size was frequency dependent to ensure the point spread function (PSF) was at least three pixels wide at zenith. For each timestep, 31 images were created - one for each fine frequency channel (the first of the 32 frequency channels was dropped due to calibration issues). The image size was kept constant for each of these 31 images in frequency, which required regridding to ensure that the RA/Dec cell size remained constant with frequency. XX and YY polarisation images were computed separately to produce a total of $\sim$76 million images for the full survey.

The method for detecting and identifying satellites in these images is the same as our work in \citet{grigg_paper3}, and a brief summary of this procedure is discussed below. For clarity, `fine channel images' refers to the 31 individual 28.9 kHz channel images, and `coarse channel images' refers to the mean stack of these 31 fine channel images to give 0.90 MHz bandwidth per image. A visual inspection of these data showed broadband and bright narrowband emission from Starlink satellites, so the coarse channel images were used to make detections. This was because the broadband emission was not always visible above the noise floor in the fine channel images, but became visible in the coarse channel images when the broadband emission was stacked.

In each coarse channel image, the predicted locations\footnote{from \url{space-track.org}} of all satellites with a perigee $<$2,000 km were calculated. At each of these predicted locations, a 2D Gaussian fit was attempted at each of these predicted locations. For a successful fit, the following conditions had to be met: (1) the elevation had to be above 20$^{\circ}$, (2) the uncertainty in both amplitude and the standard deviation of the Gaussian had to be $<$5\%, (3) the fitted centre of the Gaussian could not deviate by more than the larger of two pixels or three degrees from the predicted position, and (4) the fit could not be over the Sun or a bright radio galaxy. If the fit passed these constraints, it was labelled as a detection. Once a list of detections had been compiled for each dataset, the list was analysed to find when five or more detections closely matched the trajectory of the single pass of a satellite, and if so they were marked as an identification. These identifications are presented in Section \ref{sec:results}.

\begin{figure}[h]
    \centering
    \begin{minipage}{0.5\textwidth}
        \centering
        \begin{subfigure}{\textwidth}
            \centering
            \includegraphics[width=\textwidth]{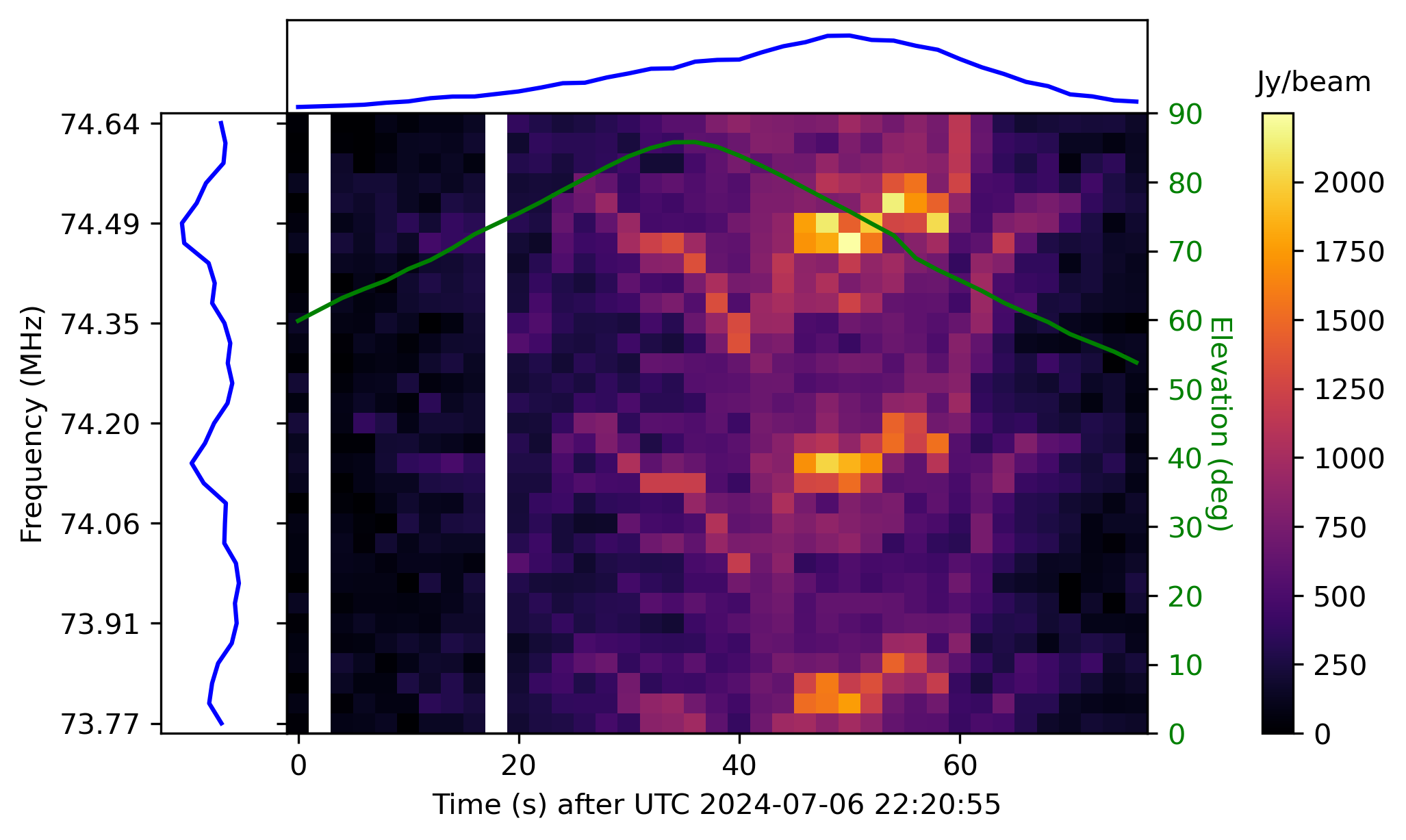}
        \end{subfigure}
        \begin{subfigure}{\textwidth}
            \centering
            \includegraphics[width=\textwidth]{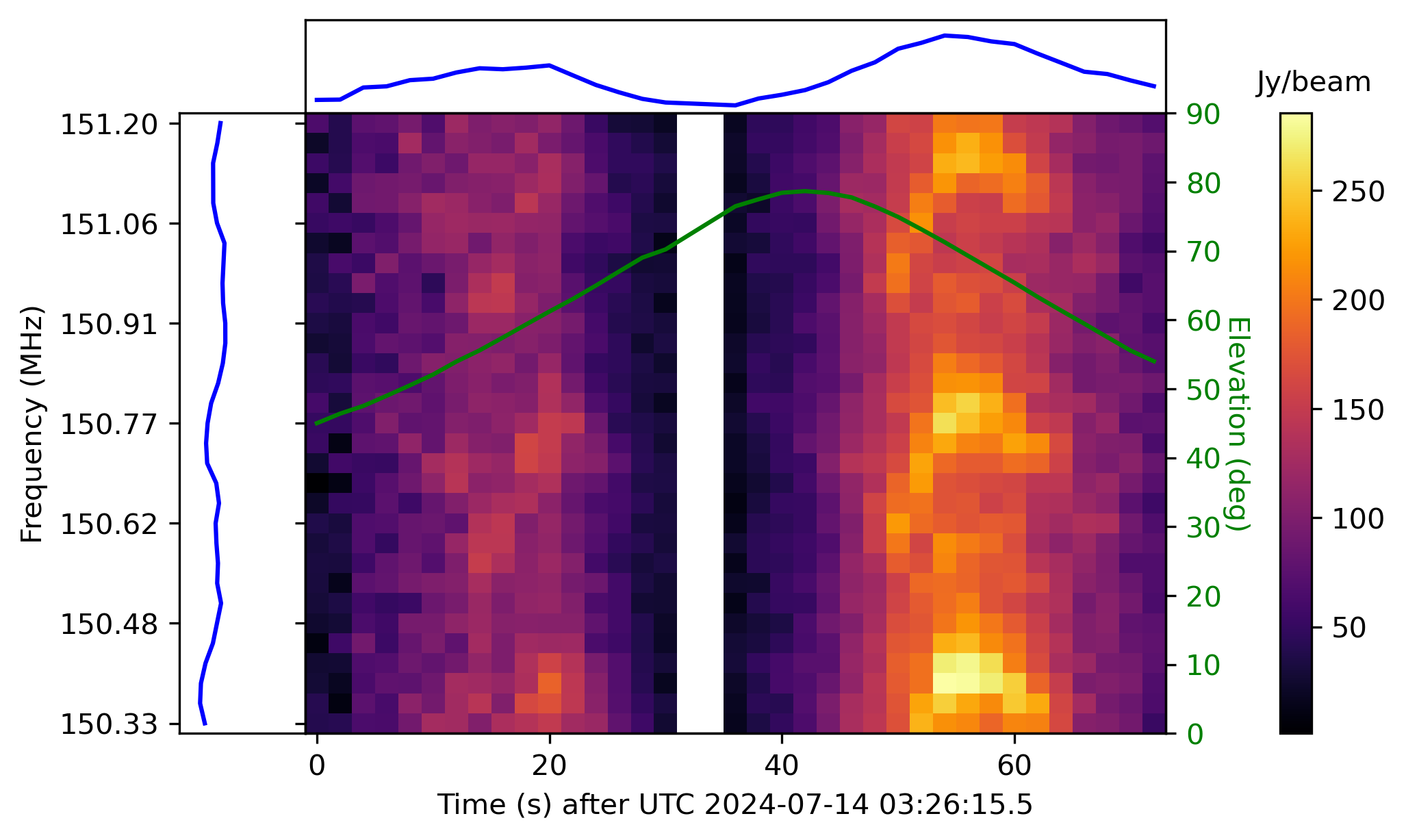}
        \end{subfigure}\\
    \end{minipage}
    \caption{\label{fig_waterfall} Waterfall plots of two Starlink satellite passes: STARLINK-31967 (NORAD 60021, v2-mini Ku, top) and STARLINK-11173 (NORAD 60192, v2-mini DTC, bottom) within two ITU protected frequencies. The amplitudes in the fine frequency channels are the pixel amplitudes at the fitted location of the satellite. The blue lines show the integrated flux density over a single frequency/time step and the green line shows the elevation of the satellite with respect to the EDA2 (90$^{\circ}$ being at the zenith). Timesteps with no values indicate that no fit could be made at the satellite's predicted location.}
\end{figure}

\begin{figure*}[h]
    \centering
    \includegraphics[width=1\textwidth]{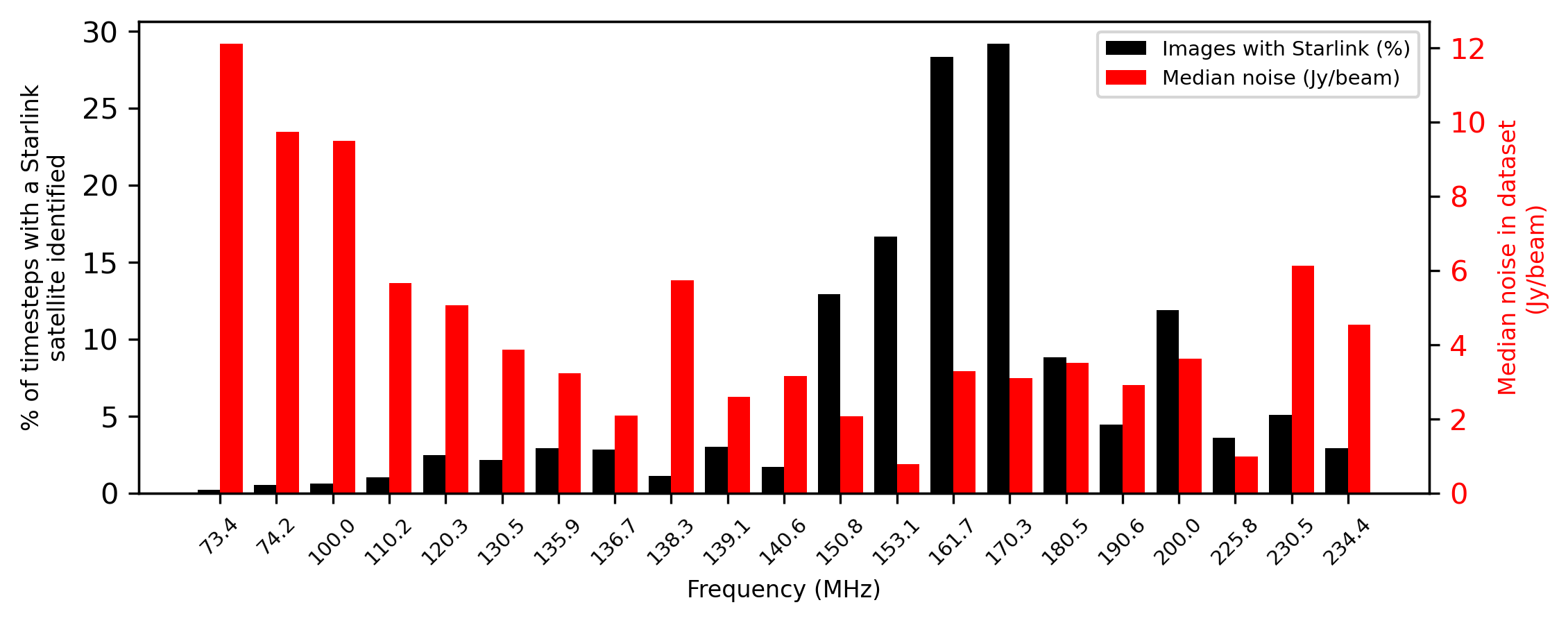}
    \caption{\label{fig_percentages} The percentage of images in each dataset with at least one Starlink satellite identified. Image noise was calculated by performing two passes of pixel excision above 3$\sigma$ and calculating the standard deviation of pixel values across a central box in the image of size 10\% of the image. This is similar to the procedure used in \citet{aegean}.}
\end{figure*}

An attempt was made to automatically classify identifications as broadband or narrowband emission, and to do this the 31 fine channel images were used. At the location of each individual identification from the coarse channel image, the pixel amplitude was extracted from all corresponding 31 fine channel images at the same timestep. This results in one column of pixels in Figure \ref{fig_waterfall}. 

To make this classification, a 1-D Gaussian with equation $A \cdot \exp\left(-\left[\frac{(\nu - \nu_{0})^{2}}{2 {\sigma}^2}\right]\right)$ was fit to these extracted fine channel amplitudes for every individual timestep, where: A is the amplitude, $\nu$ is the frequency of the input, $\nu_{0}$ is the peak frequency of the Gaussian, and $\sigma$ is the standard deviation (width) of the Gaussian and related to the full width at half maximum (FWHM) by FWHM $= 2\sigma\sqrt{2ln(2)}$. If $\sigma$ was less than 28.9 kHz (one fine channel) and the fit had a coefficient of determination ($R^{2}$; \citep{pearson}) value $>$0.95, then the emission at a single timestep was determined to likely be narrowband. If any of these criteria were not met it was assumed to be broadband. 

This left three categorisations for a single pass of a satellite: broadband, narrowband, or a combination of both. There was no evidence of Starlink satellites with two narrowband downlink frequencies within 0.926 MHz of each other in these data.

\section{Results}
\label{sec:results}

In total, 112,534 individual identifications of 1,806 unique Starlink satellites were made across the full survey. A full list of these identifications, with detailed metadata, is made available with this work and can be accessed here\footnote{\url{https://doi.org/10.5281/zenodo.15089852}}. A video of images demonstrating some identifications in the 150.78 MHz dataset can be accessed here\footnote{\url{https://youtu.be/4Fi9AAEzSxk}}. 

On the final day of data acquisition, there were 6,513 Starlink satellites in orbit, meaning that 28\% of all Starlink satellites were identified (keeping in mind not all Starlink satellites had been launched in all datasets). This comprised 54,109 and 58,425 detections in the XX and YY polarisation images, respectively. The breakdown per dataset is shown in Table \ref{table_starlink}.

It is worth noting that for an unresolved point source, the peak amplitude of a Gaussian fitted over the source's entire angular extent is a good approximation for the intensity at the centre of the source, meaning that the source's flux density is equivalent to its intensity in this case.

Figure \ref{fig_percentages} illustrates the ubiquitousness of Starlink satellites in the surveys by showing the percentage of images that contain at least one Starlink identification for each frequency. The temporal rate of occurrence of identifications peaks in the 161.7 and 170.5 MHz datasets, with almost 30\% of images containing at least one identification. In reality, these numbers are likely higher. Some of the detection parameters, such as requiring the uncertainty on the fitted value of the amplitude (A) and width of the Gaussian ($\sigma$) to be less than 5\%, are set strictly in order to limit misidentifications. More detail can be found in \citet{grigg_paper3}. Relaxing these parameters would increase the number of legitimate identifications whilst also significantly increasing the number of misidentifications. Some datasets were also heavily affected by noise, and Figure \ref{fig_percentages} shows that in general, datasets with a higher median noise have fewer images with at least a single Starlink satellite detected in them. The higher noise at lower frequencies was largely due to the Milky Way being brighter, but there were also other causes of noise such as terrestrial RFI and significant lightning in some datasets. 

\begin{figure}[h]
    \centering
    \includegraphics[width=0.45\textwidth]{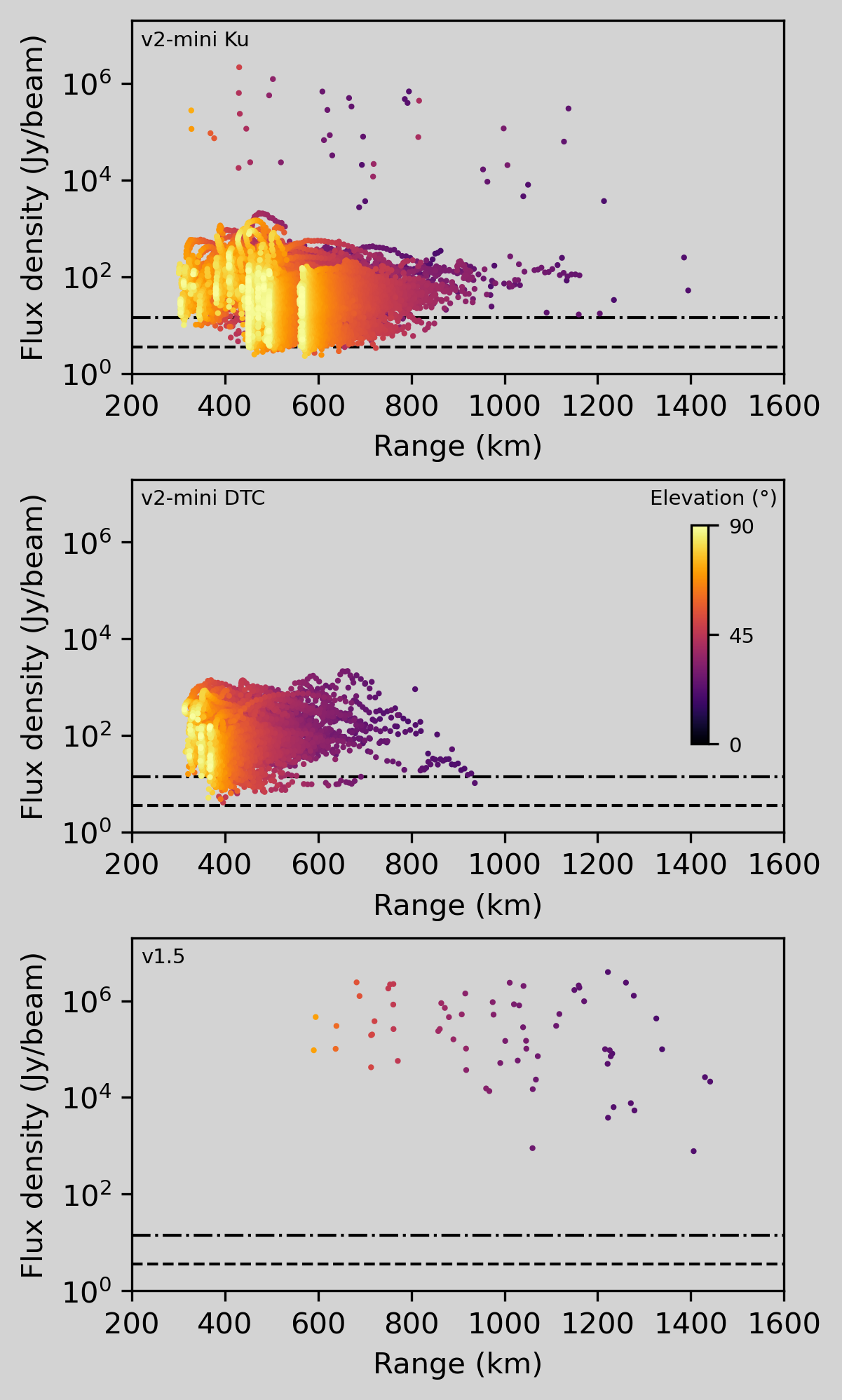}
    \caption{\label{fig_intensity_vs_range} Flux density as a function of range for the three models of detected Starlink satellites for all 54,109 XX polarisation identifications. The dotted lines show the median noise of all images in the XX datasets, with the top line being the dataset with the highest noise and the bottom line showing the median of all datasets. Image noise was calculated by performing two passes of pixel excision above 3$\sigma$ and calculating the standard deviation of pixel values across a central box in the image of size 10\% of the image. This is similar to the procedure used in \citet{aegean}.}
\end{figure}

It is also important to note that no Starlink satellites were detected in the three datasets above 320 MHz. The Sun has a higher flux density at higher frequencies, increasing image noise during the day, but none were detected at night. This could also be because the sensitivity of the antennas is significantly diminished above 300 MHz \citep{eda2}. This could still help constrain an upper frequency limit on the satellites' broadband UEMR.

It is useful to understand the total flux density due to Starlink satellites visible in the sky on average throughout these datasets. Excluding the datasets at 136.7 MHz (containing mostly intentional narrowband emission) and datasets $>320$ MHz (no identifications), the mean flux density in the sky in the frequency averaged images due to Starlink emission across both polarisations was 93 Jy/beam. This was calculated by taking the flux density of every Starlink satellite identification and dividing it by the number of timesteps and polarisations. This was as high as 312 Jy/beam at 170.3 MHz and as low as 7 Jy/beam at 225.8 MHz.

\subsection{Starlink Models}

Three models of Starlink satellite were detected, all displaying different temporal and spectral properties in their emissions. These were version 1.5 (v1.5), version 2 mini Ku-band (v2-mini Ku), and version 2 mini direct-to-cell (v2-mini DTC). No model v1.0 satellites were detected. For reference, out of the 6,513 Starlink satellites in orbit on the last day of data acquisition, 33\% were v2-mini Ku, 4\% v2-mini DTC, 44\% v1.5, and 19\% v1.0.

The majority of identifications were type v2-mini Ku, with 1,623 Starlink satellites of this model detected. This is 76\% of all v2-mini Ku satellites in orbit. Their emission displayed two distinct characteristics, as shown in Figure \ref{fig_intensity_vs_range}. The majority of identifications were broadband in nature across the full 0.926 MHz observational bandwidth. These broadband identifications are the cluster of identifications at below $\sim$1,000 Jy/beam in the top panel in Figure \ref{fig_intensity_vs_range}. This broadband emission is detected across all datasets below 320 MHz. 

The second kind of emission from v2-mini Ku satellites was recorded at a cadence of every 100 s, very similar to the behaviour of Starlink satellites reported in \citep{grigg_starlink} at 137.50 $\pm$ 0.46 MHz. In our work, we use 31 x 28.9 kHz channels, and constrain a downlink frequency of 137.05 MHz with a FWHM of 13 kHz. The highest recorded instance was 7.7 MJy/beam, orders of magnitude brighter than the broadband emission. 137.05 MHz is just outside the 137.50 $\pm$ 0.46 MHz range of the \citet{grigg_starlink} work, so it is not clear if these are emission at two different downlink frequencies or if the emission was so bright that it spilled over into the lower frequency band.

The next most common detections were of model v2-mini DTC, with 175 detected. This is 71\% of the total population of v2-mini DTC satellites. The first satellites of this model were launched on 2024-01-02. This model displayed only broadband behaviour, and no narrowband signals were detected from these satellites. Figure \ref{fig_intensity_vs_range} shows that these are only detected below a 400 km range at high elevations (coloured yellow in the Figure).

Lastly, the v1.5 model was also detected, with eight satellites identified. The v1.5 satellites were not detected as broadband signals, only the narrowband signals which pulsed every 100 s. These can be seen in the bottom panel in Figure \ref{fig_intensity_vs_range}. They appeared identical to the same narrowband signals from the v2-mini satellites with the same downlink frequency of 137.05 MHz.

\subsection{Comparison with Other Studies}

There have been four recent papers detailing the detection of Starlink satellites across SKA-Low frequencies. These are \citet{lofar}, \citet{grigg_starlink}, \citet{bassa_starlink}, and \citet{grigg_paper3}. It is interesting to compare detections between previous works to understand the evolution of the Starlink constellation.

In the first paper by \citet{lofar}, 68 Starlink satellites are detected. Of these, 12 were model v1.0 (both "Initial" and "Visorsat" models), and the other 46 were v1.5 satellites. None of these 68 satellites were detected in our work even though 55 of them were still in orbit during this survey. In \citet{lofar}, the narrowband emission at 125, 135, 143, 150, and 175 MHz does not overlap any of the frequencies in our work, with the broadband emission detected from the v1.5 satellites being $\sim$1 Jy, which is below the EDA2's sensitivity level in these observations.

\begin{figure*}[h]
    \centering
    \includegraphics[width=1\textwidth]{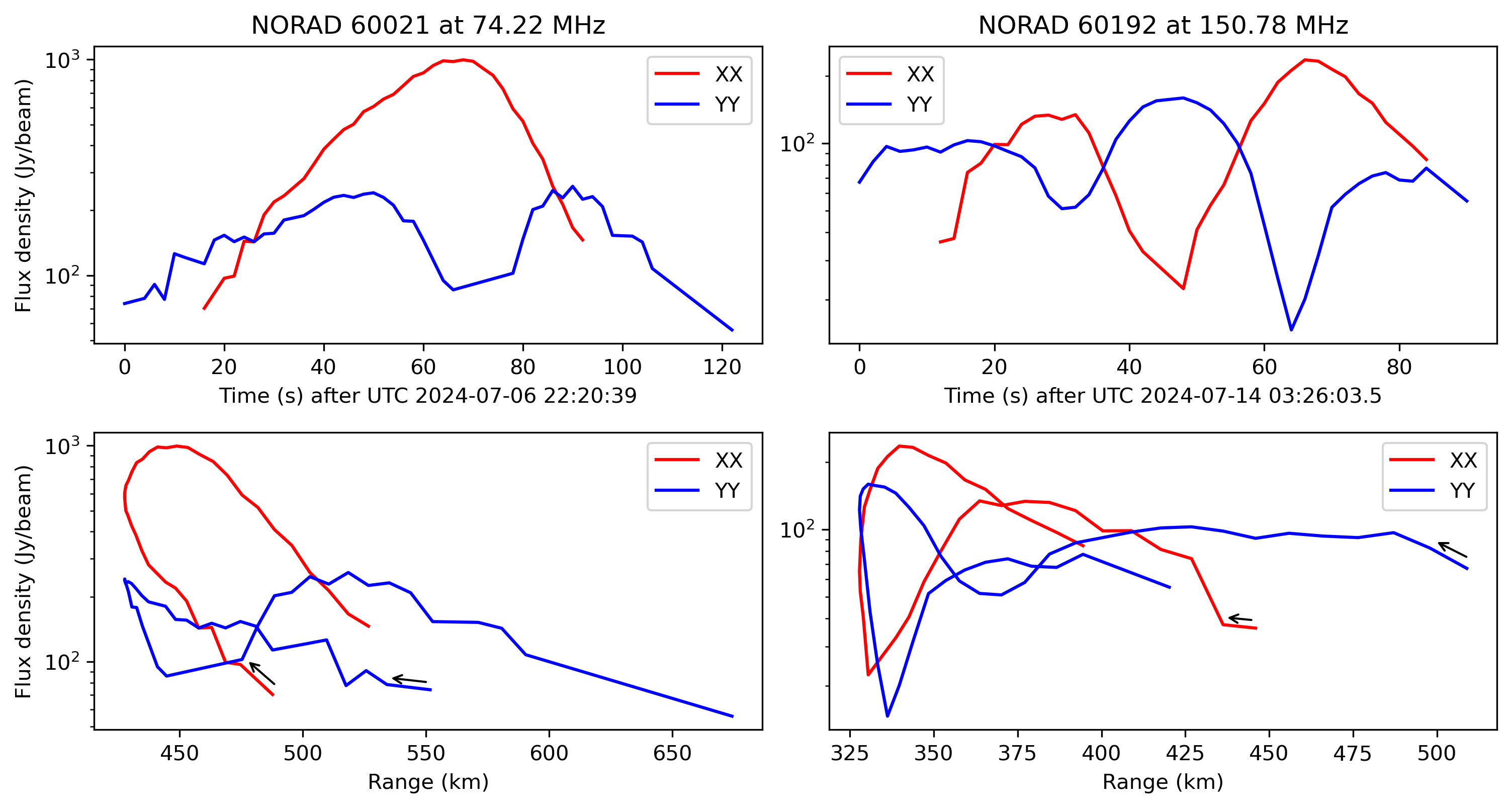}
    \caption{\label{fig_lightcurves} Flux density as a function of both time and TLE predicted range for XX and YY polarisations of the satellites STARLINK-31967 (NORAD 60021, v2-mini Ku, left) and STARLINK-11173 (NORAD 60192, v2-mini DTC, right) within two ITU protected frequency ranges.}
\end{figure*}

In the next paper by \citet{grigg_starlink}, 136 Starlink satellites are detected at 137.5 MHz, of which 20 are v2-mini Ku and 116 are v1.5. Although the paper lacks the frequency resolution required to classify the signal as either broadband or narrowband, its behaviour - pulsing once every 100 s - is the same as the narrowband detections made in this new work at 137.05 MHz. There is also a train of v1.5 and v2-mini Ku satellites seen transmitting UEMR at 159.4 MHz, although due to the close proximity of the satellites in the train it is unclear which model is the cause of this UEMR. There is no broadband signal detected at 137.5 MHz in \citet{grigg_starlink}, which is likely due to the channel being largely dominated by transmissions with a greater received flux density than the broadband emission we see in this current work. It is also interesting to note that the two bottom plots in Figure \ref{fig_lightcurves} also show the same behaviour as Figure 4 in \citet{grigg_starlink} where the satellites' flux density is modulated by something other than the range, and that the flux densities of the two polarisations are anti-correlated. 

The paper by \citet{bassa_starlink} detected 136 Starlink satellites, which were of v1.0, v1.5, v2-mini Ku, and v2-mini DTC. Of these, 47 were also detected in our work, but only the v2-mini Ku and v2-mini DTC models. This paper reports similar narrowband signals as the \citet{lofar} work. They also report a "comb" like frequency structure in the broadband emission of the v2-mini Ku and v2-mini DTC satellites. We also see this spectral structure, and additionally see that it varies with time. Figure \ref{fig_waterfall} shows an example of a v2-mini satellite at 74.22 MHz and a v2-mini DTC satellite at 150.78 MHz. \citet{bassa_starlink} note that this spectral structure is less apparent in their 110 - 188 MHz band than in their 10 - 88 MHz band. Our data also show this trend, as a higher percentage of detections at lower frequencies - especially in the 73.44/74.22 MHz datasets - show this spectral structure. This being said, there are also examples of strong spectral structure at frequencies in the 150 - 160 MHz range as shown in Figure \ref{fig_waterfall}, which are not seen in the 110 - 188 MHz observations in Appendix A in \citet{bassa_starlink}. Figure \ref{fig_waterfall} also shows that the spectral structure changes in time, with the frequency gap between peaks staying similar. This could give hints on the origin of the emission.

The last paper by \citet{grigg_paper3} detects 26 v2-mini Starlink satellites at 160.2 MHz. Although that work only uses a single 0.926 MHz frequency channel, the behaviour of the emission appears similar to the broadband UEMR detected from Starlink satellites in our current work. It is also interesting to note that 22 of these 26 were detected in the current work at frequencies ranging from 120.31 MHz to 225.78 MHz.

\subsection{Polarisation behaviour}
The lower plot in Figure \ref{fig_waterfall} shows the received flux density of the UEMR from Starlink satellites does not necessarily increase as a function of the elevation of the satellite with respect to the EDA2. In most cases, there is a sinusoidal modulation of the received signal. Interestingly, the XX and YY polarisation flux densities are anti-correlated at any given time. Figure \ref{fig_lightcurves} shows examples of this behaviour in both a v2-mini Ku and a v2-mini DTC satellite, with the flux densities of the two polarisations displaying this behaviour.

\begin{figure}[ht]
    \centering
    \includegraphics[width=0.45\textwidth]{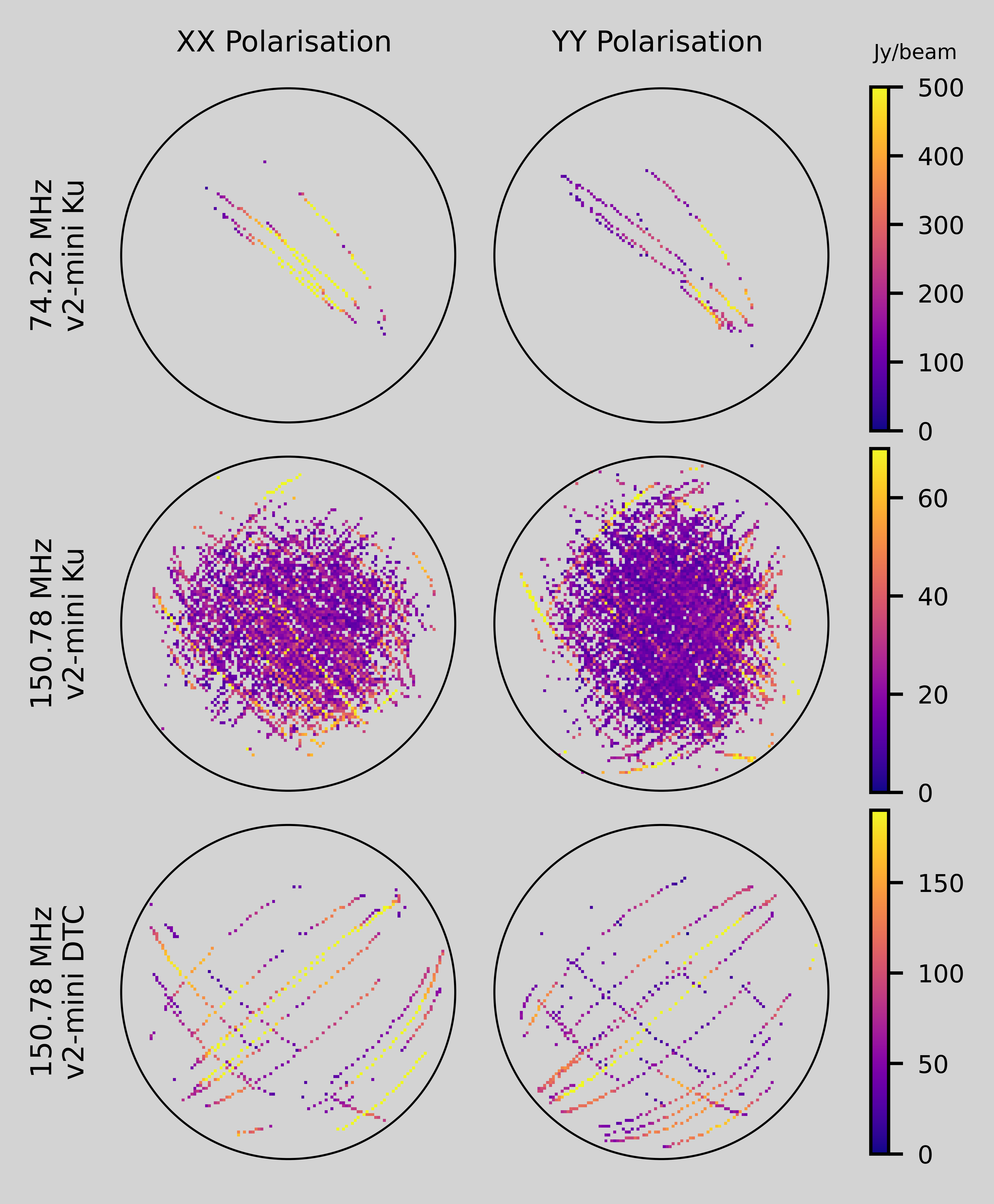}
    \caption{\label{fig_intensities_on_sky} The flux density, as a function of position on the sky, relative to the EDA2, for all Starlink satellites of a given model at a given frequency. The flux densities are from identifications of the satellites in images with all frequency channels averaged - resulting in a 0.926 MHz bandwidth. The black circle surrounding each plot is the 20$^{\circ}$ elevation isobar on the sky.}
\end{figure}

Figure \ref{fig_intensities_on_sky} shows how this effect manifests in the flux densities as a function of the satellites' motion over the sky. All three cases show the broadband UEMR of either v2-mini Ku or v2-mini DTC satellites at two frequencies. The plots highlight the differing behaviour of the two models at the two different frequencies.

\subsection{Reflections}
\label{results_reflections}
There were four examples of Starlink satellites displaying narrow-band emission at a central frequency of 99.70 MHz. This appears to be reflected terrestrial FM radio signal from a 10 kW FM transmitter in Geraldton - a city $\sim$300 km South West from the EDA2. In each pass of the four satellites, the received signal is stronger when the satellite is closest to Geraldton. Two were v2-mini Ku (NORADs 60143 and 60733) satellites, and the other two were v2-mini DTC (NORADs 60722 and 60997). From the TLE predicted ranges, three of the satellites were \textless345 km from the EDA2 at closest approach, while NORAD 59260 had a closest approach of 511 km. All satellites are still in orbit at the time of writing this paper. NORAD 60997 was launched four days before being detected. NORADs 60722 and 60733 were two different models of Starlink satellite, had similar flux densities, were detected within a few seconds of each other, and were launched two weeks before being detected. These three satellites were likely still being raised to their operational altitude. NORAD 59260, as well as being much higher in orbit, was also launched six months before being detected.\footnote{Although detections of non-Starlink satellites are out of scope for this work, the International Space Station and OSCAR 30 were both detected via reflections at the same frequency of 99.70 MHz.} \citet{lofar} also detected reflections off Starlink satellites from the GRAVES radar system at 143.05 MHz.

There was an additional identification of a v2-mini Ku satellite (NORAD 58093) at exactly 100.00 MHz. The detected signal was only visible in a single fine channel. Terrestrial FM radio transmit frequencies are spaced 200 kHz apart on odd multiples of 100 kHz, meaning this does not line up with any regular FM radio transmission frequency. The satellite is also detected at 170.31 MHz, transmitting broadband UEMR similar to other satellites. The origin of the signal at 100 MHz is currently unknown.

\section{Discussion and conclusions}
\label{sec:discussion}

This study presents the largest survey of Starlink satellites at SKA-Low frequencies to date, demonstrating the prevalence of their emission at key science and protected frequencies. The results show that Starlink satellites produce intended and unintended radio emission between 73 - 235 MHz. Figure \ref{fig_percentages} shows that a significant percentage of images - up to 29\% in some datasets - contain a detectable Starlink satellite. The dominant presence of v2-mini (especially v2-mini Ku) satellites underscores the rapid rise in Starlink satellites being launched into orbit. These findings are of increasing concern to radio astronomy.

A major finding of this study is the spectral and temporal spread of Starlink UEMR, now extremely widespread throughout the SKA-Low's observing bandwidth. Attempting to detect the EoR is a major goal for the SKA-Low, and \citet{2020MNRAS.498..265W} estimate that an EoR dataset containing as much as 1 mJy of RFI can overwhelm the EoR power spectrum integration (with additional caveats). With the caveats listed in the results, we estimate a lower limit on the mean flux density visible in the sky across all datasets and polarisations to be 93 Jy/beam for the Starlink UEMR. Although not all this energy would be visible in the power spectrum used in the EoR work, and acknowledging that Starlink satellites are not always visible in the sky, this is $\sim$5 orders of magnitude higher than the limit from the \citet{2020MNRAS.498..265W} work. With the increased sensitivity of the SKA-Low, this number will likely be even higher. This Starlink broadband UEMR alone could potentially prevent detection of the EoR.

The current ITU-R Radio Regulations do not regulate UEMR, and we observe it over frequencies allocated to downlink frequencies of other satellites. It's also highly visible in primary and secondary allocations for radio astronomy by the ITU. Our results complement those from the LOFAR telescope, demonstrating that Starlink satellites are transmitting in some of these protected frequencies. In the 73.00 - 74.60 MHz protected band, we detect 13 unique v2-mini Ku satellites, and in the 150.05 – 153.00 MHz band we detect 703 unique v2-mini satellites (684 v2-mini Ku and 19 v2-mini DTC).

The ITU's radio regulations (for Astronomy) predominantly focus on managing intentional transmissions from spacecraft within allocated frequency ranges, rather than accounting for unintended or incidental emission that electronic components or antenna systems can also produce. It has been extensively demonstrated with the EDA2 and the LOFAR that Starlink satellites produce such emission.

Currently, the specific component of the satellite that causes this unintended emission is unknown. In \citet{grigg_starlink}, SpaceX previously suggested that the emission was coming from the propulsion or avionics system and is likely produced at 50 - 200 MHz as the emission was observed as the satellites were being raised into their operational altitude. This emission has now been comprehensively detected from Starlink satellites which have reached their operational altitude. The propulsion system will likely still be engaged periodically to maintain this altitude, but the observed character of the emission is constant. The emission is also detected above 200 MHz at three frequency channels around 230 MHz. The periodic spectral structure seen in Figure \ref{fig_waterfall} and in Appendix A in \citet{bassa_starlink} gives definite clues into the source of this emission, which will be investigated in future work.

Starlink satellites - at least historically - had a different orientation when they were deployed in their parking orbit (as they are released) to when they reached their operational altitude. \footnote{\url{https://www.spacex.com/updates}}. When orbit raising, they have an "open book" orientation, where the satellite has the solar panels out flat in front of the vehicle to minimise cross-sectional drag in the direction of the thrust, transitioning to a "shark fin" orientation at operational altitude, with the phased-array antennas pointing towards Nadir. To reduce sunlight reflections off the solar panels, the satellites are, where possible, rotated during orbit raising so that the solar panels are `knife edge' towards the sun. This could lead to a favourable orientation for terrestrial radio energy to be reflected off the solar panel and back to Earth. This could be why reflections of terrestrial FM radio are detected off some Starlink satellites with a favourable geometry in both this work and \citet{lofar}. It would be interesting for further work to consider the orientation of the solar panels when reflections are detected for radio observations.

The final data products from this analysis are made available to assist in utilising the data most effectively for future regulation. We hope they can be used as a reference to current levels of unintended emissions and to be directly compared to future studies in an effort to quantify future increases or decreases in the emission. We have made available data for all identified Starlink satellites with detailed metadata (such as position, time, flux density etc.) for every identification in every image, as well as the corresponding waterfall plots for each pass (which can also be created from these data). We hope this transparency enables future work to further understand the source of this emission.

This work contributes to the ongoing dialogue on the coexistence of radio astronomy and satellite mega-constellations. The unprecedented expansion of the Starlink constellation is now significantly affecting radio astronomy observations as demonstrated in this work and others \citep{lofar, grigg_starlink, bassa_starlink, grigg_paper3}, and this work presents findings to enable policymakers to quantify changes in the behaviour of this constellation. Future mitigation of the UEMR from Starlink satellites will be key for ultra-sensitive experiments with the SKA - an example of these being the study of the Epoch of Reionisation. SpaceX has made significant progress liaising with astronomers in the optical domain, and we hope to keep this dialogue open in the radio domain.

\begin{acknowledgements}
We want to pay our respects to the traditional custodians of Inyarrimanha Ilgari Bundara, the Wajarri Yamatji people, for allowing us to conduct our science on their land. We also thank Steve Prabu for his discussions in the development of this work. Data processing took place at the Perth office of DUG Technology. Many software packages were utilised in this work, and we wish to acknowledge the developers and maintainers. The main packages were \texttt{MIRIAD} \citep{miriad}, and various Python projects such as \texttt{Astropy} \citep{astropy}, \texttt{Skyfield} \citep{skyfield}, and \texttt{Numpy} \citep{numpy}.
\end{acknowledgements}

\bibliographystyle{aa_bib}
\bibliography{bib}

\end{document}

%% file: results_table.tex
\begin{table*}[htbp]
    \centering
    \caption{\label{table_starlink} A summary of the 29 datasets presented in this work. An identification is per Starlink satellite per image. Note that unique NORAD counts are specific to each dataset.}
    \begin{tabular}{|c|c|c|c|cc|cc|}
    \hline
        \multicolumn{4}{|c|}{} & \multicolumn{2}{c|}{XX Polarisation} & \multicolumn{2}{c|}{YY Polarisation} \\ \hline
        \begin{tabular}[c]{@{}c@{}}Central\\ Frequency (MHz)\end{tabular} & 
        \begin{tabular}[c]{@{}c@{}}N$^{o}$\\ Timesteps\end{tabular} &
        \begin{tabular}[c]{@{}c@{}}Start Time\\ (UTC)\end{tabular} &
        \begin{tabular}[c]{@{}c@{}}Duration\\ (hours)\end{tabular} & 
        Identifications & 
        \begin{tabular}[c]{@{}c@{}}Unique\\ NORADs\end{tabular} & 
        Identifications & 
        \begin{tabular}[c]{@{}c@{}}Unique\\ NORADs\end{tabular} \\ \hline
        
        73.44  & 42784 & 20240627 01251900 & 11.9 & 67    & 2   & 55    & 3   \\
        74.22  & 89800 & 20240706 00301900 & 25.0 & 152   & 5   & 192   & 5   \\
        150.78 & 86984 & 20240714 02131738 & 24.2 & 3885  & 192 & 5301  & 252 \\
        153.13 & 91840 & 20240721 01284544 & 25.6 & 5611  & 273 & 7024  & 321 \\
        225.78 & 86966 & 20240728 01271636 & 24.2 & 1077  & 55  & 807   & 45  \\
        230.47 & 86986 & 20240810 00451612 & 24.2 & 2192  & 77  & 512   & 33  \\
        234.38 & 72528 & 20240811 01015710 & 22.4 & 203   & 14  & 553   & 35  \\
        322.66 & 86982 & 20240813 09451629 & 24.2 & 0     & 0   & 0     & 0   \\
        135.94 & 93696 & 20240814 11185735 & 27.4 & 1050  & 46  & 1241  & 56  \\
        136.72 & 62086 & 20240815 12101662 & 24.0 & 1046  & 40  & 1049  & 44  \\
        138.28 & 87238 & 20240823 03445919 & 25.9 & 339   & 13  & 282   & 15  \\
        139.06 & 87434 & 20240824 05453009 & 24.4 & 921   & 41  & 926   & 41  \\
        100.00 & 86990 & 20240909 07483413 & 24.2 & 161   & 9   & 162   & 10  \\
        150.78 & 86990 & 20240910 11060245 & 24.2 & 3165  & 161 & 4528  & 215 \\
        200.00 & 86990 & 20240912 05121446 & 24.2 & 3543  & 172 & 3903  & 166 \\
        120.31 & 86990 & 20240914 02322986 & 24.2 & 719   & 25  & 825   & 27  \\
        180.47 & 86988 & 20240915 02554858 & 24.2 & 2733  & 141 & 2534  & 123 \\
        110.16 & 79570 & 20240918 08250726 & 24.3 & 150   & 7   & 335   & 14  \\
        130.47 & 86990 & 20240919 23542941 & 24.2 & 780   & 22  & 914   & 27  \\
        170.31 & 86988 & 20240921 01120222 & 24.2 & 12185 & 522 & 11318 & 483 \\
        140.63 & 86990 & 20240922 02225646 & 24.2 & 433   & 14  & 507   & 24  \\
        161.72 & 86990 & 20240923 02545465 & 24.2 & 10381 & 469 & 11679 & 519 \\
        190.63 & 86990 & 20240924 05055809 & 24.2 & 1453  & 66  & 1438  & 64  \\
        73.44  & 86990 & 20240925 08454398 & 24.2 & 38    & 3   & 39    & 3   \\
        325.00 & 86978 & 20241018 09230333 & 24.2 & 0     & 0   & 0     & 0   \\
        328.13 & 86978 & 20241020 05382177 & 24.2 & 0     & 0   & 0     & 0   \\
        136.72 & 86984 & 20241023 09245832 & 24.2 & 705   & 34  & 879   & 38  \\
        234.38 & 86986 & 20241025 05291905 & 24.2 & 772   & 42  & 1097  & 53  \\
        110.16 & 86986 & 20241026 09232772 & 24.2 & 348   & 12  & 325   & 12  \\ 
        \hline
        Total & 2446692 & - & 694.2 & 54109 & - & 58425 & - \\ \hline
    \end{tabular}
\end{table*}